\documentclass[structabstract]{aa}  
\usepackage{graphicx}
\usepackage{xcolor}
\usepackage{txfonts}

\renewcommand*{\Psi}{\varPsi}
\renewcommand*{\Omega}{\varOmega}

\newcommand{\kep}{{{\it Kepler }}}

\newcommand{\rhr}{r_\mathrm{hr}}

\def\be{\begin{equation}}
\def\ee{\end{equation}}    
\def\ba{\begin{eqnarray}}
\def\ea{\end{eqnarray}}

\begin{document}

\title{Asteroseismic stellar activity relations}
\author{A. Bonanno\inst{1},
E. Corsaro\inst{2,1},
C. Karoff\inst{3,4}
          }
\offprints{Alfio Bonanno\\ \email{alfio.bonanno@inaf.it}}

\institute{
INAF - Osservatorio Astrofisico di Catania, via S. Sofia, 78, 95123 Catania, Italy
\and Instituut voor Sterrenkunde, KU Leuven, Celestijnenlaan 200D, B-3001 Leuven, Belgium
\and Stellar Astrophysics Centre, Department of Physics and Astronomy, Aarhus University, Ny Munkegade 120, DK-8000 Aarhus C, Denmark
\and Department of Geoscience, Aarhus University, H{\o}egh-Guldbergs Gade 2, 8000 Aarhus C, Denmark
}

\date{}
\abstract{ 
In asteroseismology an important diagnostic of the evolutionary status of a star is the small frequency 
separation which is sensitive to the gradient of the mean molecular weight in the stellar interior.
It is thus interesting to discuss the classical age-activity relations in terms of this quantity. 
Moreover, as the photospheric magnetic field tends to suppress the amplitudes of acoustic oscillations, it is important to quantify the importance of this effect by considering various activity indicators.} 
{We propose a new class of age-activity relations that connects the Mt. Wilson $S$ index and the average scatter in the light curve with the small frequency separation and the amplitude of the p-mode oscillations.}
{We used a Bayesian inference to compute the posterior probability of various empirical laws for a sample of 19 solar-like active stars observed by the {\it Kepler} telescope.}
{ We demonstrate the presence of a clear correlation between the Mt. Wilson $S$ index and the relative age of the stars as indicated by the small frequency separation, as well as an anti-correlation between the $S$ index and the oscillation amplitudes. We argue that the average activity level of the stars shows a stronger correlation with the small frequency separation than with the absolute age that is often considered in the literature.} 
{The phenomenological laws discovered in this paper have the potential to become new important diagnostics to link stellar evolution theory with the dynamics of 
global magnetic fields. In particular we argue that the relation between the Mt. Wilson $S$ index  and the oscillation amplitudes is in good agreement with the findings  of direct numerical 
simulations of magneto-convection.} 
%
\titlerunning{Asteroseismic stellar activity relations}
      \authorrunning{A. Bonanno et al.}
\maketitle
%
\section{Introduction}
\label{sec:intro}
It is well known that in an active star the average activity level decreases as the star evolves along the main sequence because it reduces 
its rotation rate \citep{skum72}.  
The observed correlation between X-ray luminosity and rotation rate in late-type stars \citep{palla} can in fact be explained in terms of a 
dynamo action, although the precise nature of the underlying dynamo model is still a subject of debate \citep{wright}.
Most of the attempts to interpret the correlation between the rotation rate, age, and the amplitude of the activity cycles
in a unifying picture have been hampered by serious theoretical difficulties \citep{noyes84,saar,vite}.

The core of the problem is that
during various stages of the star's lifetime the type of dynamo responsible for sustaining the star's global magnetic field 
can change drastically. Young, fast-rotating stars approaching the zero-age main-sequence 
are characterized by a strong dynamo action, a possible nonlinear combination of   $\alpha\Omega$ or $\alpha^2$ mechanisms, 
producing irregular activity cycles with  large photospheric flux contrast \citep{boso,bozo}. On the contrary, main-sequence stars with extended convective zones 
can display regular activity cycles because of the presence of a tachocline, the most likely location for the $\alpha$-effect according to \cite{parker93}.

The high-quality photometric data from the  {\it Kepler} space telescope \citep{borucki10} has opened new possibilities in the investigation of these issues. In \cite{chaplin11} it was argued  that the steep fall in the observed fraction of detected stars with solar-like oscillations has a stellar explanation.
Although it was not possible to precisely quantify this effect in terms of chromospheric flux, it is conceivable to assume that 
the presence of activity suppresses the amplitude of oscillations, which hampers the possibility of detecting the stellar oscillations. 
On the other hand, recent asteroseismic studies  \citep{karoff13,gar14,mathur14} have 
further clarified the classical relations between stellar age, rotation, and various activity proxies, generally 
confirming the basic fact that the level of activity is anti-correlated with the star's age.
Interestingly enough, the analysis  of \cite{rere} has instead confirmed that stellar differential rotation is only weakly dependent on the rotational period. 

An important insight into the above questions has recently been provided by \cite{mittag}, where a 
careful study of the evolutionary status of the Mt. Wilson project stars \citep{bado}
has led  the authors to conclude that stellar activity in main-sequence stars should decline with the relative 
main-sequence age, i.e., with the evolutionary status of the star, rather than with the absolute age \citep[see also][]{remo}. 

The are two aims to this work. First, we show that there is a direct correlation between the Mt. Wilson $S$ index chromospheric activity indicator (hereafter $S_{\mathrm{MW}}$) and the small frequency separation $\delta\nu_{02}=\nu_{n,l=0}-\nu_{n-1,l+2}$.
The latter is a measure of the He fraction in the stellar core \citep{ae10,bopa}, i.e.,
a precise indicator of the evolutionary status of the star, thus confirming the idea expressed in \cite{mittag}. 
Second, we also show that there is a clear anti-correlation between $S_{\mathrm{MW}}$ and the amplitude of maximum power of the p-mode oscillation envelope, $A_\mathrm{max}$. It is then clear that the mechanism observed to suppress the p-mode amplitudes in the Sun \citep{chaplin00} is also present in other Sun-like stars.

In particular we make use of a Bayesian inference to precisely and accurately quantify the functional relations expressing our empirical laws. 

\section{New asteroseismic relations} 
\label{sec:scaling_relations}
Our targets are part of the {\it Sounding Stellar Cycles With Kepler} program \citep{karoff09,karoff13}, which combines high-precision photometric observations
from \kep with ground based observations from the Nordic Optical Telescope (NOT). In particular, we shall focus on a sample of $N_\mathrm{stars} = 19$ main-sequence solar-like stars described and analyzed by \cite{karoff13}, for which asteroseismic measurements for the large frequency separation, $\Delta\nu$, and the small frequency separation, $\delta\nu_{02}$, are made  available as computed from the frequencies measured by \cite{App12}, according to the method described in \cite{Bedding04}. In addition, we enrich the set of asteroseismic measurements by including the oscillation amplitudes at maximum power $A_\mathrm{max}$, obtained by \cite{huber11}, which we found available for 18 targets of our sample, and by computing another index, known as $\rhr$ or the \emph{range}, related to the average scatter in the \kep light curve and derived according to the method described by \cite{basri10} by using long cadence Q6 data \citep[see also][for more details]{chaplin11}. All the data presented here are listed in Table~\ref{tab:data}.

\begin{table*}
\caption{Values for the asteroseismic parameters and the range index, together with 1-$\sigma$ error bars, used for the relations presented in Sect.~\ref{sec:scaling_relations}.}             
\centering                         
\begin{tabular}{l r r r r}       
\hline\hline
\\[-8pt]         
KIC ID & \multicolumn{1}{c}{$\Delta\nu$} & \multicolumn{1}{c}{$\delta\nu_{02}$} & \multicolumn{1}{c}{$A_\mathrm{max}$} & \multicolumn{1}{c}{$r_\mathrm{hr}$}\\ 
 & \multicolumn{1}{c}{($\mu$Hz)} & \multicolumn{1}{c}{($\mu$Hz)} & \multicolumn{1}{c}{(ppm)} & \multicolumn{1}{c}{(ppt)}\\
\hline
\\[-8pt]
01435467 & $70.43\pm1.10$ & $4.924\pm1.512$ & $6.768\pm0.440$ & $0.471\pm0.099$\\[2pt]
02837475 & $75.66\pm1.49$ & $6.787\pm2.661$ & $6.787\pm0.310$ & $0.193\pm0.111$\\[2pt]
03733735 & $91.85\pm1.67$ & $9.058\pm3.808$ & $5.174\pm0.351$ & $0.222\pm0.017$\\[2pt]
04914923 & $88.54\pm1.55$ & $5.082\pm0.980$ & \multicolumn{1}{c}{$\dots$} & \multicolumn{1}{c}{$\dots$}\\[2pt]
06116048 & $100.50\pm0.59$ & $6.224\pm0.949$ & $5.975\pm0.340$ & $0.290\pm0.118$\\[2pt]
06603624 & $109.90\pm0.46$ & $5.387\pm0.423$ & $5.666\pm0.470$ & $0.358\pm0.037$\\[2pt]
06933899 & $71.85\pm0.48$ & $5.100\pm0.546$ & $8.448\pm0.503$ & $0.318\pm0.018$\\[2pt]
07206837 & $78.84\pm2.38$ & $6.494\pm3.726$ & $7.003\pm0.480$ & $0.504\pm0.150$\\[2pt]
08006161 & $149.19\pm0.34$ & $10.140\pm0.688$ & $2.906\pm0.265$ & $0.650\pm0.189$\\[2pt]
08379927 & $119.94\pm0.63$ & $10.655\pm1.208$ & $3.568\pm0.165$ & $1.682\pm0.577$\\[2pt]
08694723 & $75.22\pm0.65$ & $6.182\pm1.032$ & $7.939\pm0.359$ & $0.286\pm0.031$\\[2pt]
09098294 & $109.27\pm1.33$ & $5.903\pm1.900$ & $5.997\pm0.593$ & $0.586\pm0.188$\\[2pt]
09139151 & $117.09\pm1.87$ & $10.484\pm2.631$ & $4.624\pm0.448$ & $0.602\pm0.091$\\[2pt]
09139163 & $81.23\pm0.87$ & $6.662\pm1.397$ & $5.911\pm0.290$ & $0.452\pm0.061$\\[2pt]
10454113 & $105.10\pm1.25$ & $9.171\pm1.410$ & $4.285\pm0.308$ & $0.557\pm0.038$\\[2pt]
11244118 & $71.33\pm0.53$ & $5.208\pm0.489$ & $8.953\pm0.650$ & $0.577\pm0.341$\\[2pt]
11253226 & $76.92\pm2.56$ & $4.811\pm4.126$ & $6.615\pm0.312$ & $0.470\pm0.187$\\[2pt]
12009504 & $87.96\pm0.58$ & $6.454\pm0.954$ & $6.923\pm0.381$ & $0.319\pm0.041$\\[2pt]
12258514 & $74.69\pm0.54$ & $4.854\pm0.876$ & $7.000\pm0.303$ & $0.456\pm0.225$\\[2pt]
\hline                                
\end{tabular}
\label{tab:data}
\end{table*}

In the context of this work we have investigated new empirical asteroseismic relations that involve the activity proxies $S_{\mathrm{MW}}$ and $\rhr$ and the asteroseismic parameters $\delta\nu_{02}$ and $A_\mathrm{max}$. Similarly to the analysis done by \cite{corsaro13} for the amplitude scaling relations in asteroseismology, we have adopted the following equations in logarithmic form: 
\begin{equation}
\ln S_{\mathrm{MW}} = a_1 + b_1 \ln \delta\nu_{02} \, ,
\label{eq:S_d02}
\end{equation}
\begin{equation}
\ln \rhr = a_2 + b_2 \ln \delta\nu_{02} \, ,
\label{eq:r_d02}
\end{equation}
\begin{equation}
\ln S_{\mathrm{MW}} = a_3 + b_3 \ln A_\mathrm{max} \, ,
\label{eq:S_A}
\end{equation}
\begin{equation}
\ln \rhr = a_4 + b_4 \ln A_\mathrm{max} \, .
\label{eq:r_A}
\end{equation}
The free parameters $\left( a_i, b_i \right)$ for $i$ ranging from 1 to 4, are estimated by means of a Bayesian analysis. For this purpose, we exploited the log-likelihood introduced by \cite{corsaro13}, assuming that the $S$ indices are log-normally distributed, which we write here as
\begin{equation}
\Lambda \left( a_i, b_i \right) = \Lambda_0 - \frac{1}{2} \sum^{N_\mathrm{stars}}_{j=1} \left[ \frac{\Delta_j \left( a_i, b_i \right) }{\widetilde{\sigma}_j \left( b_i \right)} \right]^2
\end{equation}
with
\begin{equation}
\Lambda_0 = - \sum^{N_\mathrm{stars}}_{j=1} \ln \sqrt {2 \pi \widetilde{\sigma}_j \left( b_i \right)} \, 
\end{equation}
and
\begin{equation}
\Delta_j \left( a_i, b_i \right) = \ln S_{\mathrm{MW},j}^\mathrm{obs} - \ln S_{\mathrm{MW},j}^\mathrm{pred} \left( a_i, b_i \right) \, ,
\end{equation}
namely the difference between the measured $S$ index provided by \cite{karoff13} and the one predicted by the empirical relations for the $j$-th star. Uniform priors are adopted for all the free parameters. Still following \cite{corsaro13}, we point out that the relative uncertainties $\widetilde{\sigma}_j \left( b_i \right)$ for each target are obtained according to the formal expression of the relations investigated, hence yielding
\begin{equation}
\widetilde{\sigma}_j^2 \left( b_1 \right) = \left( \frac{\sigma_{S_{\mathrm{MW},j}}}{S_{\mathrm{MW},j}} \right)^2 + b^2_1 \left( \frac{\sigma_{\delta\nu_{02,j}}}{\delta\nu_{02,j}} \right)^2
\end{equation}
and similarly
\begin{equation}
\widetilde{\sigma}_j^2 \left( b_3 \right) = \left( \frac{\sigma_{S_{\mathrm{MW},j}}}{S_{\mathrm{MW},j}} \right)^2 + b^2_3 \left( \frac{\sigma_{A_{\mathrm{max},j}}}{A_{\mathrm{max},j}} \right)^2
\end{equation}
for Eq.~(\ref{eq:S_d02}) and Eq.~(\ref{eq:S_A}), respectively, with $\sigma_{S_{\mathrm{MW}}}$, $\sigma_{\delta\nu_{02}}$, and $\sigma_{A_{\mathrm{max}}}$ being the standard uncertainties coming from the dataset adopted. This is done in order to take into account the indeterminacy on both coordinates. A perfectly analogous treatment is applied to Eq.~(\ref{eq:r_d02}) and Eq.~(\ref{eq:r_A}) by simply replacing $S_\mathrm{MW}$ with $\rhr$.

To provide additional arguments in support of the relation between stellar activity and relative age of the star, we also consider in our analysis a relation between the $S$ index and the absolute stellar age (hereafter Age for simplicity) and another between the $S$ index and the large frequency separation of $p$ modes ($\Delta\nu$) known to be sensitive to the evolution of the star as it probes the mean stellar density \citep{ulrich86}. In particular, we exploit the stellar ages derived by \cite{karoff13} by means of  asteroseismic modeling, and therefore apply a Bayesian inference along the same lines described previously in this section to the empirical laws
\begin{equation}
\ln S_\mathrm{MW} = a_5 + b_5 \ln \mbox{Age}
\label{eq:S_Age}
\end{equation}
and
\begin{equation}
\ln S_\mathrm{MW} = a_6 + b_6 \ln \Delta\nu \, .
\label{eq:S_Dnu}
\end{equation}

\subsection{Results}
\begin{figure}
   \centering
   \includegraphics[width=8.2cm]{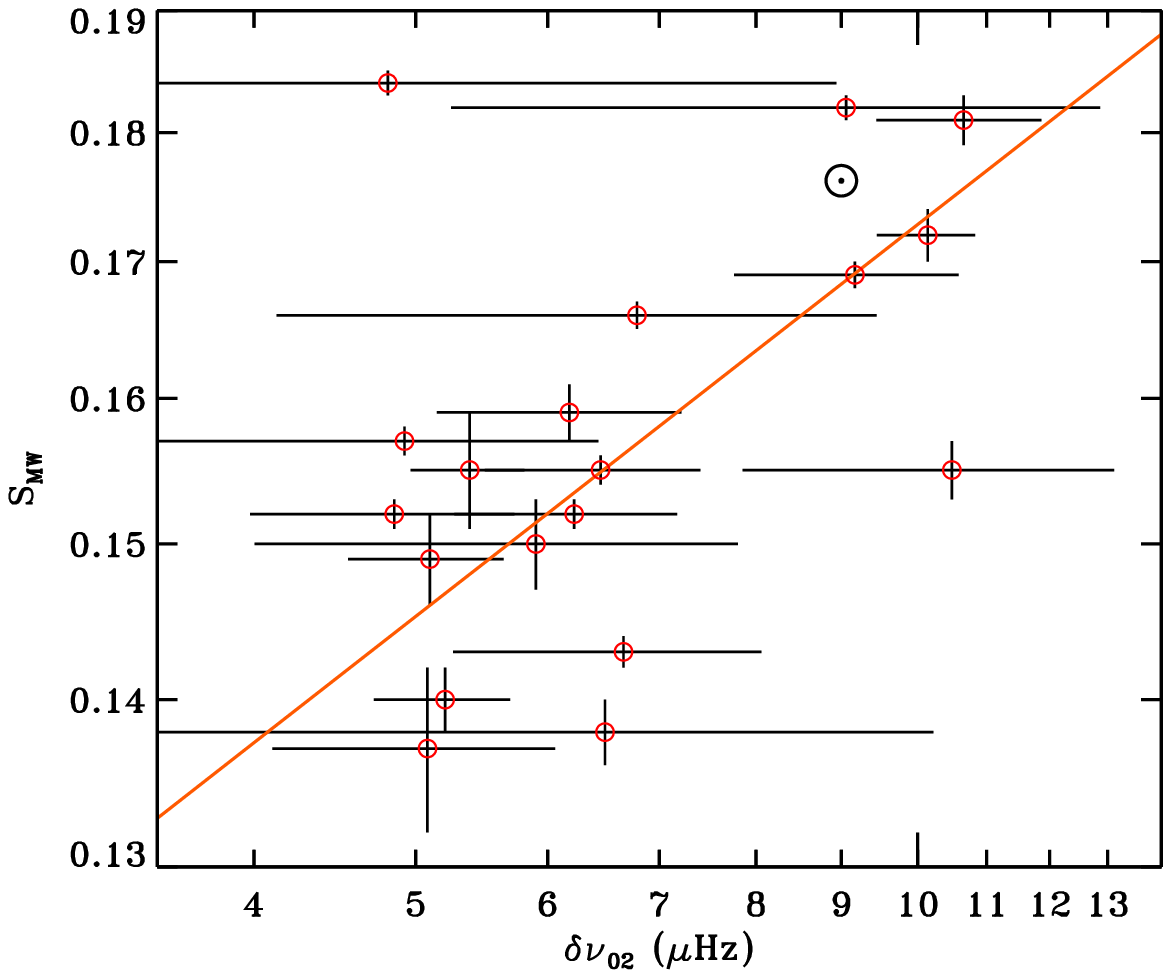}
    \includegraphics[width=8.2cm]{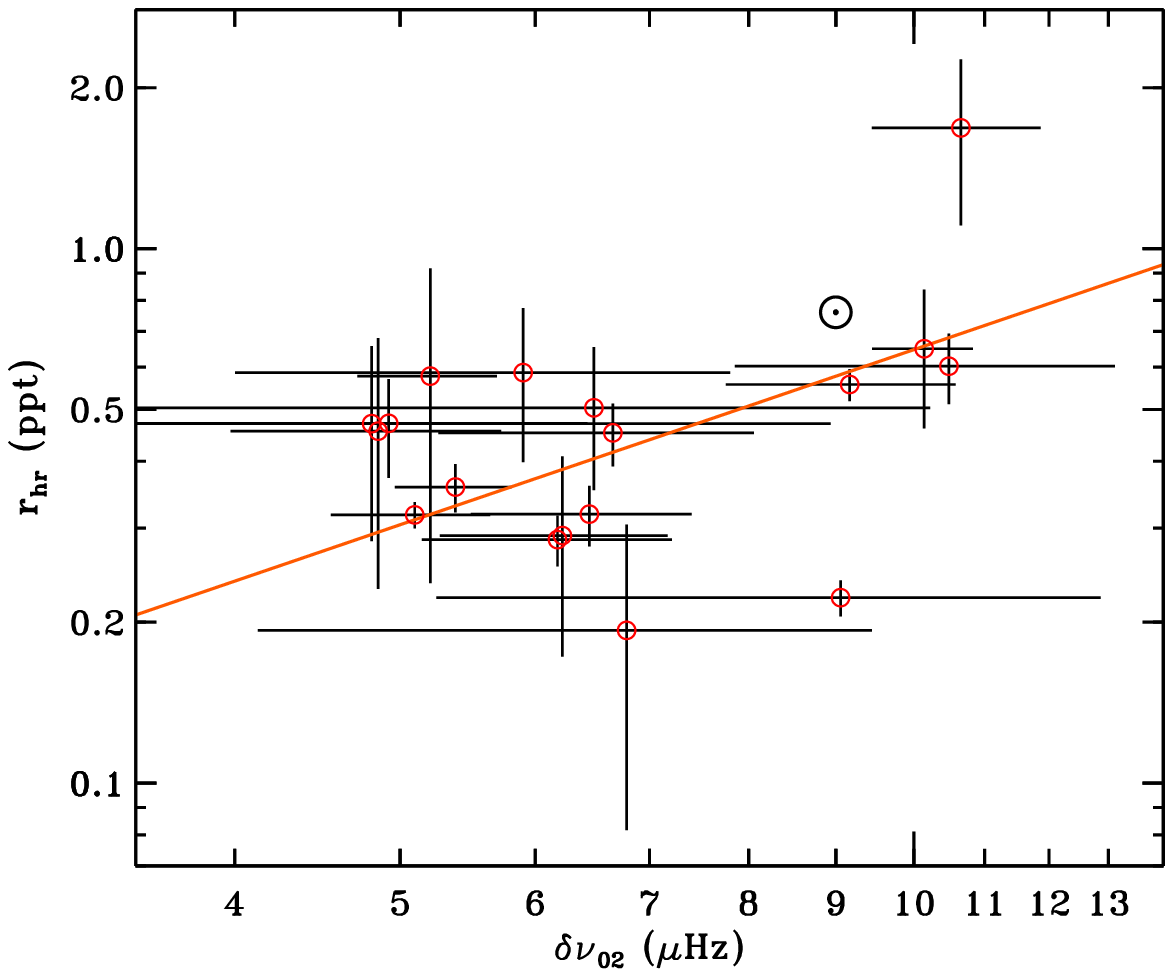}
      \caption{Log-log scale plot of the $S$ index (top) and range index (bottom) as a function of the small frequency separation $\delta\nu_{02}$. The orange solid line indicates the resulting Bayesian fits to Eq.~(\ref{eq:S_d02}) and Eq.~(\ref{eq:r_d02}), respectively, while the open symbols are the measured quantities, with corresponding error bars overlaid on both axes. The Sun symbol for the range index corresponds to the value adopted by \cite{basri10}}
    \label{fig:S_d02}
\end{figure}

\begin{figure}
   \centering
   \includegraphics[width=8.2cm]{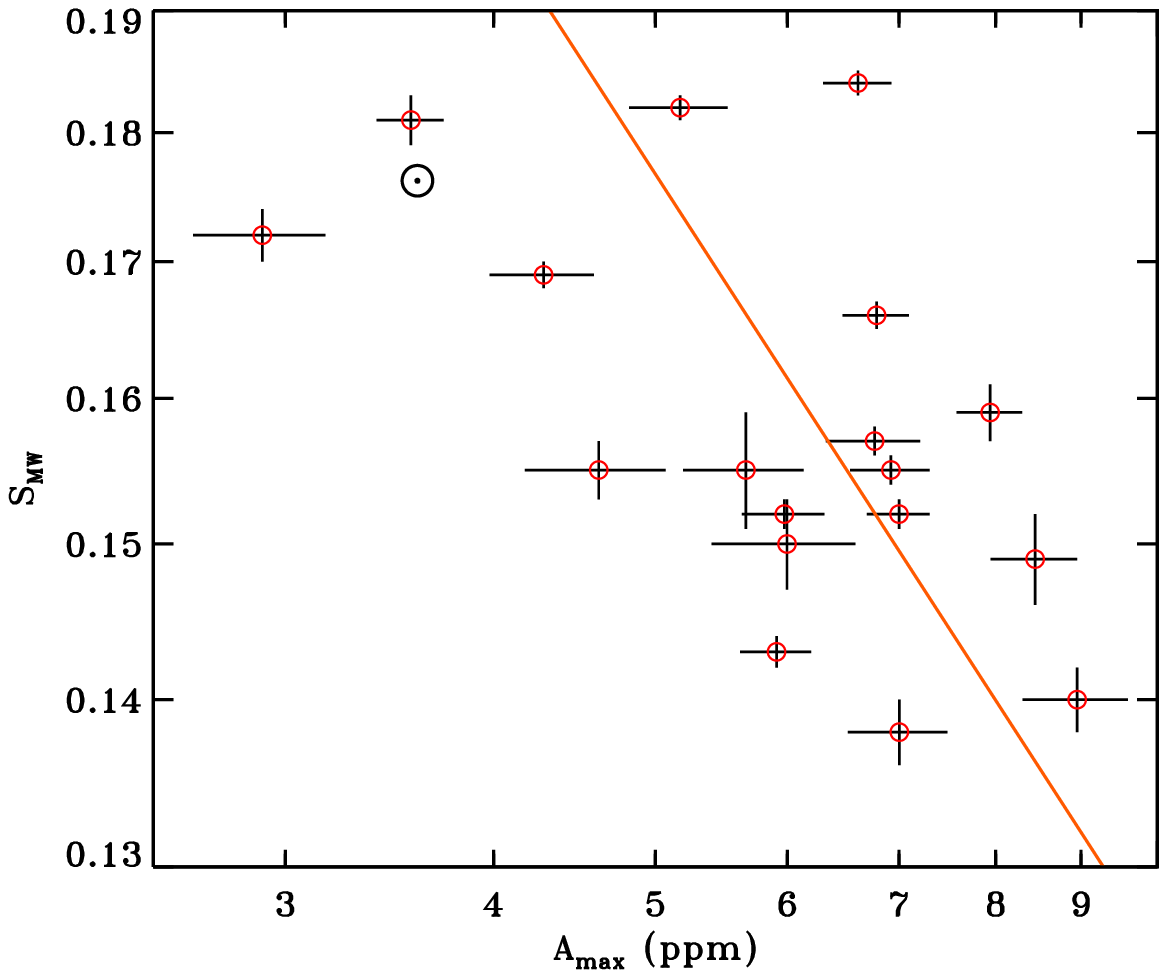}
   \includegraphics[width=8.2cm]{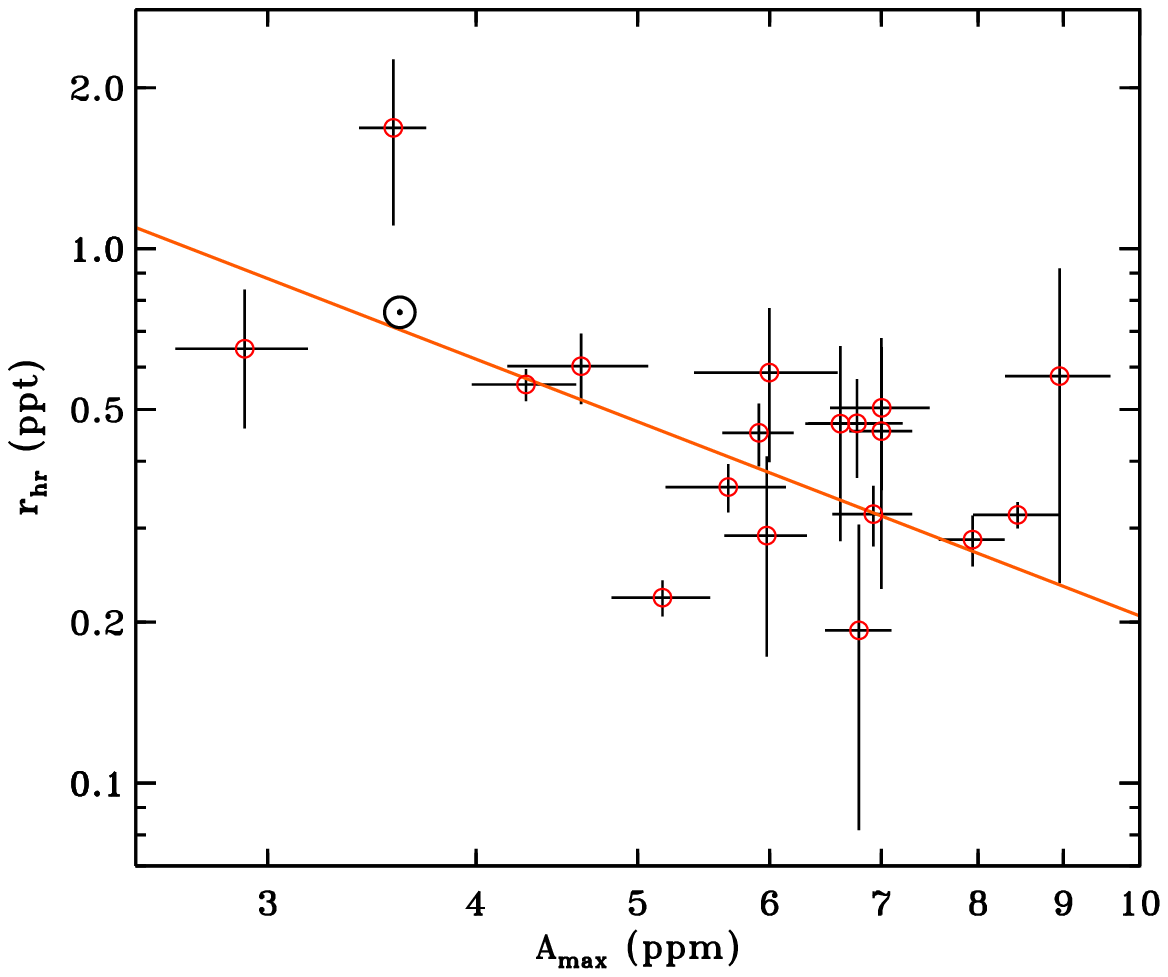}
      \caption{Same description as in Fig.~\ref{fig:S_d02}, but in the case of the relation relative to $S_\mathrm{MW}$ (top) and $\rhr$ (bottom) with $A_\mathrm{max}$, given by Eq.~(\ref{eq:S_A}) and Eq.~(\ref{eq:r_A}), respectively.}
    \label{fig:S_A}
\end{figure}

\begin{figure}
   \centering
   \includegraphics[width=8.2cm]{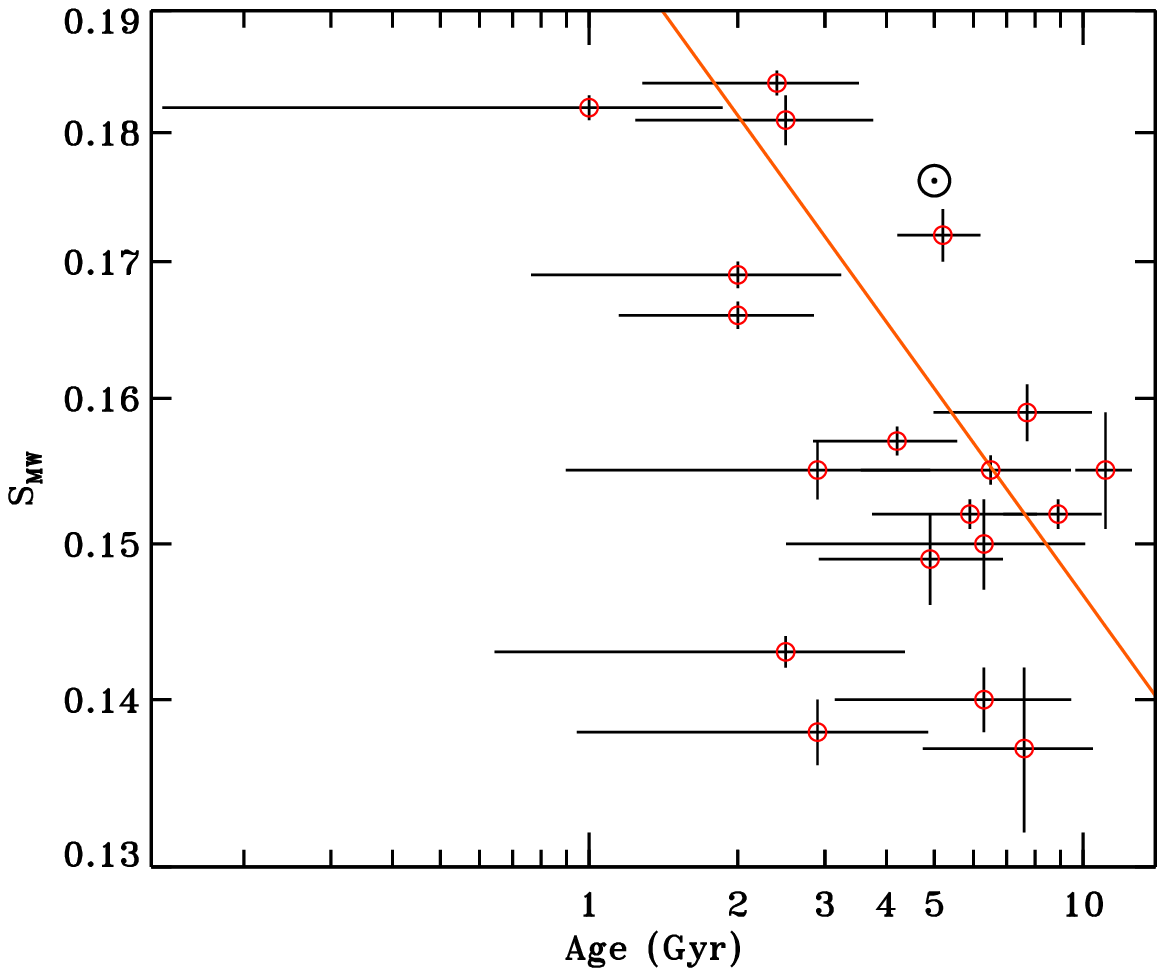}
   \includegraphics[width=8.2cm]{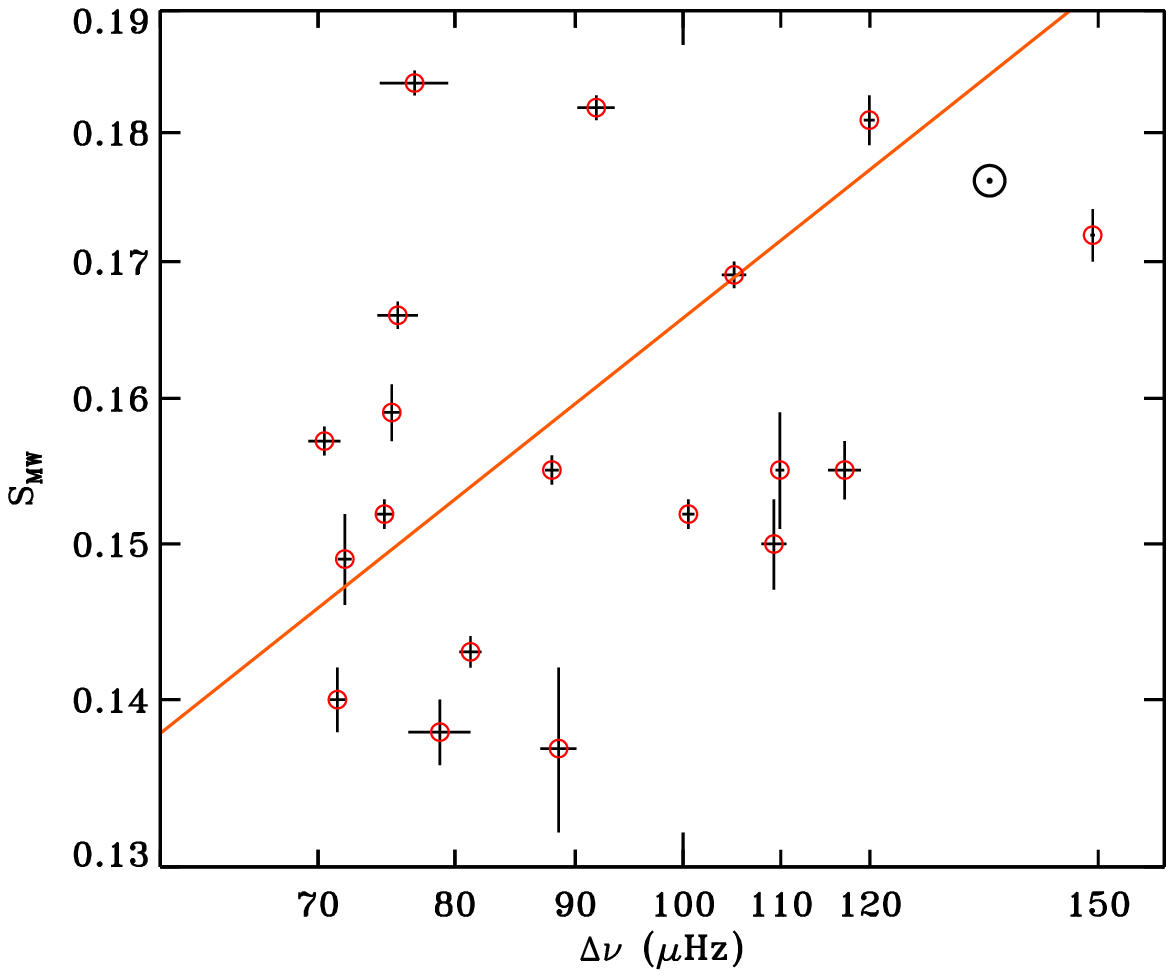}
      \caption{Same description as in Fig.~\ref{fig:S_d02}, but in this case for the relations relative to $S_\mathrm{MW}$ with $Age$ (top) and $\Delta\nu$ (bottom) given by Eq.~(\ref{eq:S_Age}) and Eq.~(\ref{eq:S_Dnu}).}
    \label{fig:S_Age}
\end{figure}

The results of the Bayesian parameter estimation are listed in Table~\ref{tab:results}, while the fitted relations are shown in Fig.~\ref{fig:S_d02} for Eq.~(\ref{eq:S_d02}) and Eq.~(\ref{eq:r_d02}), in Fig.~\ref{fig:S_A} for Eq.~(\ref{eq:S_A}) and Eq.~(\ref{eq:r_A}), and in Fig.~\ref{fig:S_Age} for Eq.~(\ref{eq:S_Age}) and Eq.~(\ref{eq:S_Dnu}).

Given the number of stars used in the inference, the coefficients of the relations for the activity index $S_\mathrm{MW}$ appear to be quite well constrained (on average around 7\,\%), ensuring enough confidence on the detected trends, even in the case of $\delta\nu_{02}$ despite its large error bars (up to $25$\,\%). For the case of $\rhr$ instead, the constraining level is much weaker (up to 34\,\%), showing that for the sample analyzed in this work the $S$ index enables us to obtain a more precise and reliable calibration of the asteroseismic activity relations by a factor of $\sim5$. This result relies on both the much larger relative error bars of the range index measurements as compared to those of the $S$ index (on average $\sim26$\,\% and $\sim1$\,\% for $\rhr$ and $S_\mathrm{MW}$, respectively) and on the different dispersion of the data points relative to the fit, as one can derive by comparing the weighted standard deviation of the residuals of the fit (on average $\sim$0.30 and $\sim$0.08 for $\rhr$ and $S_\mathrm{MW}$, respectively). 

However, apart from the significance of the fits for the two different indices, by looking at the results shown in this work we can clearly see that the stellar activity indicated by both $\rhr$ and $S_\mathrm{MW}$ increases towards larger values of $\delta\nu_{02}$ (with a slope $b_1 \simeq 0.25$ for $S_\mathrm{MW}$ and $b_3 \simeq 1.09$ for $\rhr$). When comparing this result with that obtained through the relation $S_\mathrm{MW}$-Age, we find that $S_\mathrm{MW}$ correlates with the absolute stellar age with residuals dispersed twice as much as the case $S_\mathrm{MW}$-$\delta\nu_{02}$, and that the relation $S_\mathrm{MW}$-Age also has parameter estimates with relative error bars about two times larger than for $S_\mathrm{MW}$-$\delta\nu_{02}$, together with a slope $b_5 \simeq 0.5 \, b_1$, proving that the dependence of stellar activity is weaker with the absolute age of the star. Finally, the relation $S_\mathrm{MW}$-$\Delta\nu$ gives residuals that are on average about two times more dispersed than for $S_\mathrm{MW}$-$\delta\nu_{02}$, although for the dataset used the relative error bars on $\Delta\nu$ are about 20 times smaller than for $\delta\nu_{02}$. In other words, this shows that the measured activity level of the stars in our sample is significantly more sensitive to their relative main-sequence age than to the absolute age, thus validating the argument exposed by \cite{mittag}.  

For the case of the p-mode oscillation amplitude $A_\mathrm{max}$, we clearly see that for both the $S$ and range indices, higher activity levels correspond to lower oscillation amplitudes, in agreement with the expected suppression of the oscillation signal shown by \cite{chaplin11}, which we quantify here with a slope $b_2 \simeq -0.50$ for $S_\mathrm{MW}$ and $b_4 \simeq -1.21$ for $\rhr$. 

Finally, we point out that we have tested the relations presented in Sect.~\ref{sec:scaling_relations} with the activity index $\log \langle \Delta \mathcal{F}_\mathrm{Ca} \rangle$ as provided by \cite{karoff13}, thus confirming the same trends found in the case of the $S$ index.

\begin{table}
\caption{Mean values of the free parameters (offsets $a_i$ and slopes $b_i$) indicated by Eqs. (\ref{eq:S_d02}) to (\ref{eq:r_A}) and Eqs.~(\ref{eq:S_Age}) and (\ref{eq:S_Dnu}), with corresponding 68.3\,\% Bayesian credible intervals.}             
\centering                         
\begin{tabular}{l r r}       
\hline\hline
\\[-8pt]         
Empirical relation & Offset ($a_i$) & Slope ($b_i$)\\    
\hline
\\[-8pt]
  $S_\mathrm{MW}$-$\delta\nu_{02}$ & $-2.33^{+0.06}_{-0.04}$ & $0.25^{+0.02}_{-0.03}$\\[2pt]
  $\rhr$-$\delta\nu_{02}$ & $-2.94^{+0.49}_{-0.31}$ & $1.09^{+0.17}_{-0.27}$\\[2pt]
  $S_\mathrm{MW}$-$A_\mathrm{max}$ & $-0.93^{+0.06}_{-0.08}$ & $-0.50^{+0.04}_{-0.03}$\\[2pt]
  $\rhr$-$A_\mathrm{max}$ & $1.20^{+0.33}_{-0.41}$ & $-1.21^{+0.22}_{-0.18}$\\[2pt]
  $S_\mathrm{MW}$-Age & $-1.62^{+0.03}_{-0.04}$ & $-0.13^{+0.02}_{-0.01}$\\[2pt]
  $S_\mathrm{MW}$-$\Delta\nu$ & $-3.46^{+0.06}_{-0.04}$ & $0.36^{+0.01}_{-0.01}$\\[2pt]
\hline                                
\end{tabular}
\label{tab:results}
\end{table}

\subsection{Correlating S and range indices}
\begin{figure}
   \centering
   \includegraphics[width=8.8cm]{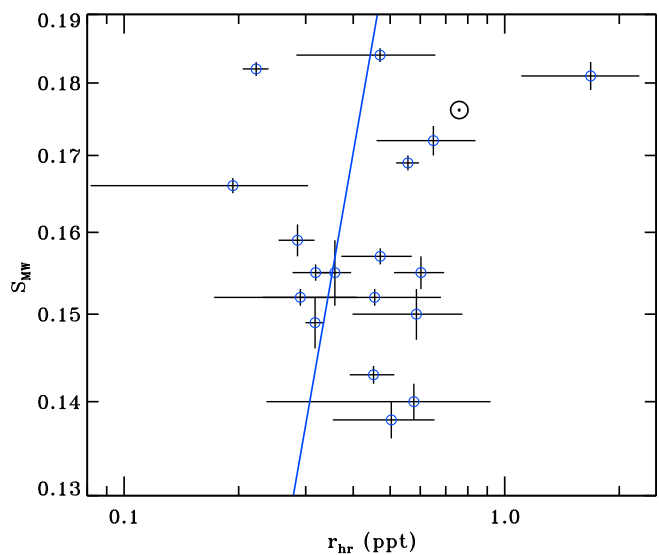}
      \caption{Same description as in Fig.~\ref{fig:S_d02}, but in the case of the relation $S_\mathrm{MW}$-$\rhr$, given by Eq.~(\ref{eq:S_r}).}
    \label{fig:S_r}
\end{figure}

In the previous section we have shown that both the $S$ and the range indices correlate similarly to the asteroseismic parameters $\delta\nu_{02}$ and $A_\mathrm{max}$. However, the range index has not yet been proven to correlate directly with the stellar activity of the star. In order to test the reliability of the range index to probe the stellar activity level we also performed the Bayesian inference on the relation
\begin{equation}
\ln S_\mathrm{MW} = \alpha + \beta \ln \rhr \, ,
\label{eq:S_r}
\end{equation}
hence estimating the parameters $(\alpha, \beta)$ by adopting once again similar arguments to those already used for Eqs.  (\ref{eq:S_d02}) to (\ref{eq:r_A}).
The result of the parameter estimation is shown in Fig.~\ref{fig:S_r}, where we can observe a direct correlation between the two indices, as already expected from the previous results shown in this work. However, the data points appear dispersed, with residuals of about 30\,\%, which together with the mean offset $\alpha = -1.09^{+0.10}_{-0.18}$ and slope $\beta = 0.75^{+0.10}_{-0.18}$ (very steep variation in the $S$ index) of Eq.~(\ref{eq:S_r}) show that for the given sample of stars, $S_\mathrm{MW}$ and $\rhr$ are not strongly correlated.

\section{Discussion and conclusions}
Can we explain the phenomenological law given by Eq.~(\ref{eq:S_A}) in terms of basic physical processes occurring at the chromospheric and photospheric level of the star?  
In the case of the Sun it has been shown that the $S$ index is roughly proportional to the local field strength $B$ \citep{sku75}. It is thus obvious to assume that a  
relation of the type $S_\mathrm{MW} \propto B$ holds to a good approximation 
for the solar-type stars considered in this study. On the other hand, direct numerical simulation of magneto-convection \citep{jac08} have proven that the power of acoustic 
oscillations is inversely proportional to the local field strength (see in particular their Fig. 4). This is in agreement with our findings for which $A_\mathrm{max} \propto 1/S^2_\mathrm{MW} \propto 1/ B^2$.
We think that this result is interesting because it opens up the possibility of obtaining important information on the local structure of the chromospheric field
by means of asteroseismology.  

It is also important to stress that, 
while most of the studies on age-activity-rotation present in the literature have focussed on 
clusters younger than about 500 Myrs, in this work we used asteroseismology to address the age-activity relation in main-sequence and subgiant stars
demonstrating the occurrence of a clear correlation between 
the small frequency separation of p-mode oscillations $\delta\nu_{02}$  and the activity index $S_\mathrm{MW}$ in a sample of 19 solar-like stars. 
Our findings suggest that a more suitable candidate with which to study the age-activity relation is $\delta\nu_{02}$  instead of the absolute age derived through asteroseismic modeling, 
which, unfortunately, is the least constrained quantity in asteroseismic studies \citep{bome}. 
We have also shown that the suppression of the p-modes energy in the pulsational spectrum due to the presence of a photospheric field is efficient, and leads to 
a rapid decrease in the p-mode amplitudes in very active stars.
Although a clear limitation of our study is the impossibility of extending this analysis to very young, fast rotating stars with strong magnetic fields, 
we argue that the idea proposed in \cite{mittag} is essentially correct.

Our testing of a possible connection between the two indices $S_\mathrm{MW}$ and $\rhr$ has shown that the very steep variation observed, together with the large dispersion of the data points, do not provide any evidence for a strong correlation. We therefore conclude that for our sample of stars the two indices are not equally sensitive to their activity level, $S_\mathrm{MW}$ being the more sensitive one, although this result would require a larger number of targets with $S$ index measurements available in order to be confirmed and extended to a wider range of stellar properties. 
Moreover, the use of the  $S_\mathrm{MW}$ index instead of $\rhr$, has a direct physical interpretation as explained above.

It is important to further investigate our new phenomenological relations using larger stellar samples and we hope to address this question in the near future.

\begin{acknowledgements}
The research leading to these results has received funding from the European
Research Council under the European Community's Seventh Framework Programme
(FP7/2007--2013) ERC grant agreement n$^\circ$227224 (PROSPERITY), 
from the Fund for Scientific Research of Flanders (G.0728.11), 
and from the Belgian federal science policy office (C90291 Gaia-DPAC). CK
acknowledges support from the Villum foundation. Funding for the Stellar
Astrophysics Centre is provided by The Danish National Research Foundation
(grant agreement No.: DNRF106).
\end{acknowledgements}

\end{document}